\documentclass[aps,prb,floatfix,twocolumn,superscriptaddress]{revtex4}
\usepackage[pdftex]{graphicx}
\usepackage{amsmath,amssymb,subfigure,color,verbatim,array}

\providecommand{\mysection}[1]{ \textit{#1} }
\providecommand{\todo}[1]{ }
\providecommand{\todo}[1]{ \red{#1} }

\providecommand{\rrr}{\mathbf{r}}

\providecommand{\calG}{\mathcal{G}}

\providecommand{\vare}{\varepsilon}

\providecommand{\vp}{\varphi}

\providecommand{\percent}{\%{}}

\providecommand{\up}{\uparrow}
\providecommand{\dn}{\downarrow}

\providecommand{\cccc}{c^{\phantom\dag}}
\providecommand{\cdag}{c^\dag}

\providecommand{\half}{\tfrac{1}{2}}

\providecommand{\mean}[1]{\left\langle #1 \right\rangle}

\providecommand{\elabel}[1]{\label{e:#1}}
\providecommand{\eref}[1]{Eq.~\eqref{e:#1}}

\providecommand{\red}[1]{{\color{red}{#1}\color{black}}}

\begin{document}
\title{Origin of Excess Low Energy States \\ in a Disordered Superconductor
in a Zeeman Field}
\author{Y.~L.~Loh}
\author{N.~Trivedi}
\affiliation{Department of Physics, The Ohio State University, 191 W Woodruff Avenue, Columbus, OH  43210}
\author{Y.~M.~Xiong}
\author{P.~W.~Adams}
\affiliation{Department of Physics and Astronomy, Louisiana State
University, Baton Rouge,
Louisiana 70803, USA}
\author{G.~Catelani}
\affiliation{Department of Physics, Yale University, 217 Prospect
Street, New Haven, CT 06520}

\date{\today} 
\begin{abstract}
Tunneling density of states measurements of disordered superconducting (SC) Al films in
high Zeeman fields reveal a significant population of subgap states which cannot be explained by standard BCS theory.
We provide a natural explanation of these excess states in terms of a 
novel disordered Larkin-Ovchinnikov (dLO) phase that occurs near the
spin-paramagnetic transition at the Chandrasekhar-Clogston critical field.
The dLO superconductor is characterized by a pairing amplitude that changes sign at domain walls.
These domain walls carry magnetization and support Andreev bound states that lead to distinct spectral signatures at low energy.
\end{abstract}
\maketitle

A central theme in condensed matter physics is the quest for new states of matter with unusual
arrangements of interacting electrons, spins, and atoms. 
The interplay between superconductivity and magnetism is an especially rich source of interesting physics that gives rise to various types of exotic superconductors such as cuprates, pnictides, ruthenates, and heavy-fermion materials
\cite{cupratesreview2006,heavyfermionreview1984}.
There is also, however, the possibility of exotic superconductivity of a different type, 
which arises when a conventional BCS superconductor at low temperature is subjected to an external Zeeman field.
In the simplest scenario, the superconductor undergoes a
first-order transition into a polarized normal Fermi liquid
\cite{chandrasekhar1962,clogston1962} when the Zeeman splitting becomes
of the order of the superconducting gap $\Delta_0$ at the Chandrasekhar-Clogston critical field  $\mu_B H_{CC}\approx {\Delta_0}/{\sqrt{2} }$.
However, nature has a more intriguing way of resolving the tussle:
the electrons can self-organize into a novel intermediate state known as a
Fulde-Ferrell-Larkin-Ovchinnikov (FFLO) state near $H_{CC}$.
\cite{fulde1964,larkin1964,machida1984,burkhardt1994,yoshida2007,loh2010}
An FFLO state 
consists of regions of positive and negative pairing amplitude
separated by domain walls where the magnetization is piled up;
it can be thought of as an ``electronic liquid crystal,''
an example of emergent microscale phase separation.
Interest in FFLO physics crosses traditional boundaries between condensed matter, cold atomic gases\cite{radzihovskyReview2010}, quantum chromodynamics\cite{casalbuoniReview2004}, nuclear physics, and astrophysics\cite{alford2001}, and there is currently an intense effort to search for FFLO phases in superconductors as well as in cold atoms\cite{liao2010}.

Hitherto, only thermodynamic signatures of the FFLO phase
have been reported, and these have been limited to a few
layered organic superconductors and the heavy fermion material CeCoIn$_5$~\cite{radovan2003,koutroulakis2010,yanase2009}.
The realization of FFLO in traditional superconducting systems has been hampered by its sensitivity to disorder and spin-orbit scattering.  
Notwithstanding these issues, we show that even in the presence of disorder, where the fully coherent FFLO phase is suppressed, spectroscopic manifestations of FFLO fluctuations are readily observable.


	\begin{figure*}[!htb]
	\subfigure[]{
		\label{GVdata}
		\includegraphics[width=0.9\columnwidth]{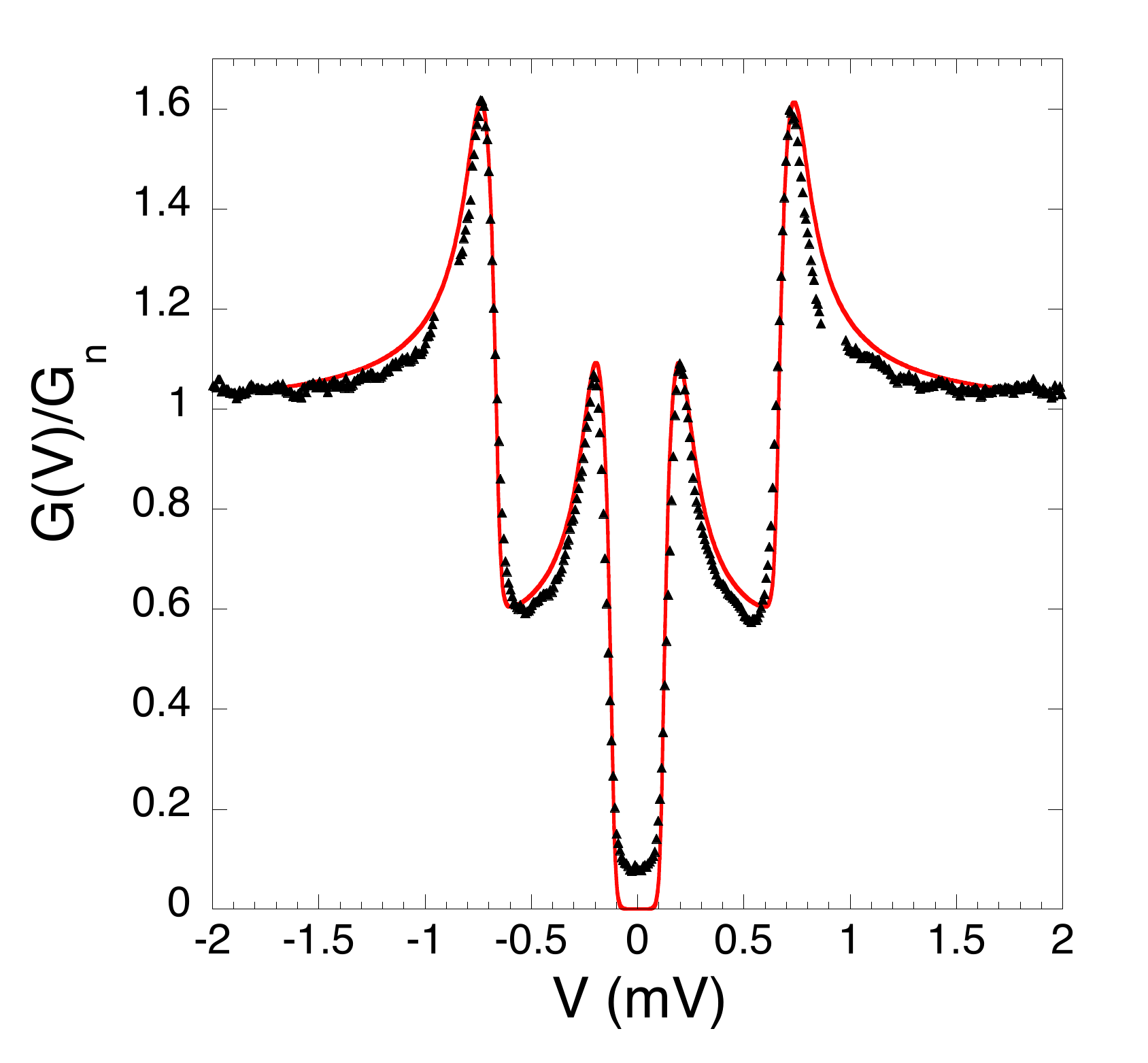}
	}
	\subfigure[]{
		\label{GHdata}
		\includegraphics[width=0.9\columnwidth]{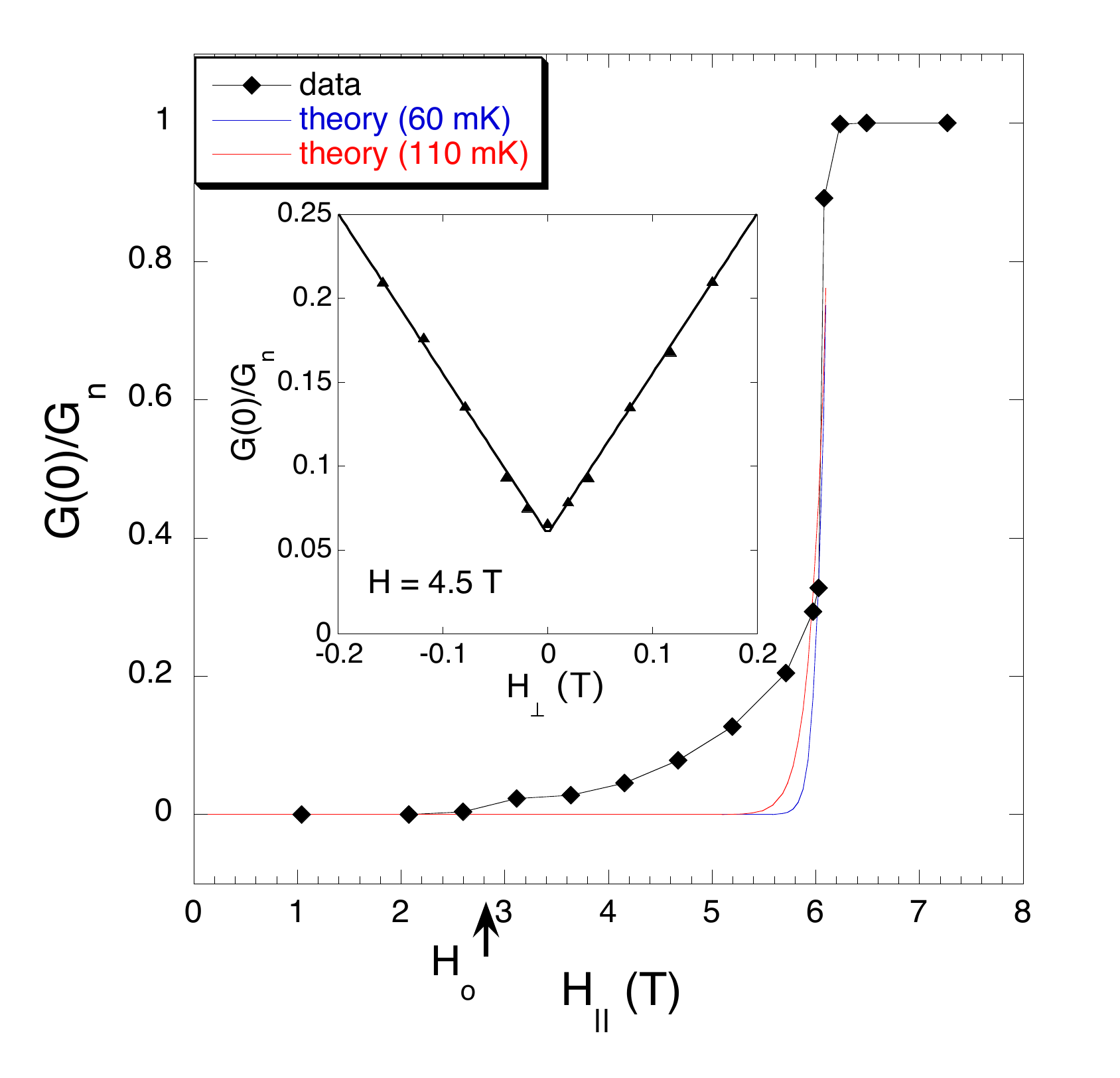}
	}
	\caption{
	\label{Gdata}
	\vspace{-0.0cm}
\small{
(a) 
Tunneling conductance $G(V)$ normalized by normal state conductance $G_n\sim(1~\rm{k}\Omega)^{-1}$ for a 24 \AA~superconducting Al film in a 4.75 T parallel field at 100 mK
(symbols=experiment, curve=homogeneous theory).
(b) 
Zero-bias tunneling conductance $G(0)$ at 60 mK as a function of parallel field $H$.
Between 
$H_0 \sim 2.8 ~\mathrm{T}$
and 
$H_{c\parallel} \sim 6.1 ~\mathrm{T}$,
the homogeneous theory (blue curve) significantly underestimates the number of states near the Fermi energy, and even when the temperature is artificially increased (red curve) it is unable to describe the broad tail in $G(0)$.
We ascribe the discrepancy to a disordered LO phase.
(Inset) Tunnel conductance as a function of $H_{\perp}=4.5\sin(\theta)$ where $\theta$ is the tilt angle $\theta$.  The solid lines are a linear least-squares fit to the data.
The sharp V-shaped minimum allows us to accurately determine parallel alignment.
	}}
	\end{figure*}

\mysection{Main results:} 
We present density of states (DoS) calculations based on a disordered attractive Hubbard model, along with low-temperature tunneling DoS
measurements on ultra-thin Al films.
We show that, contrary to popular belief, FFLO physics is not
completely washed out by disorder.
In fact, over a significant range of Zeeman fields we find a disordered
Larkin-Ovchinnikov (dLO) state characterized by bound states in domain walls and low-energy spectral weight, which provides a natural explanation of the 
experimental anomalies.\cite{adams2004} 
Our calculations self-consistently account for the disorder and allow the pairing amplitude to adjust to the disorder profile. The novel dLO phase is robust to variations in field and disorder, and imprints a unique signature in the low-energy DoS within the superconducting gap.

   \begin{figure}[!h]
   \includegraphics[width=0.9\columnwidth]{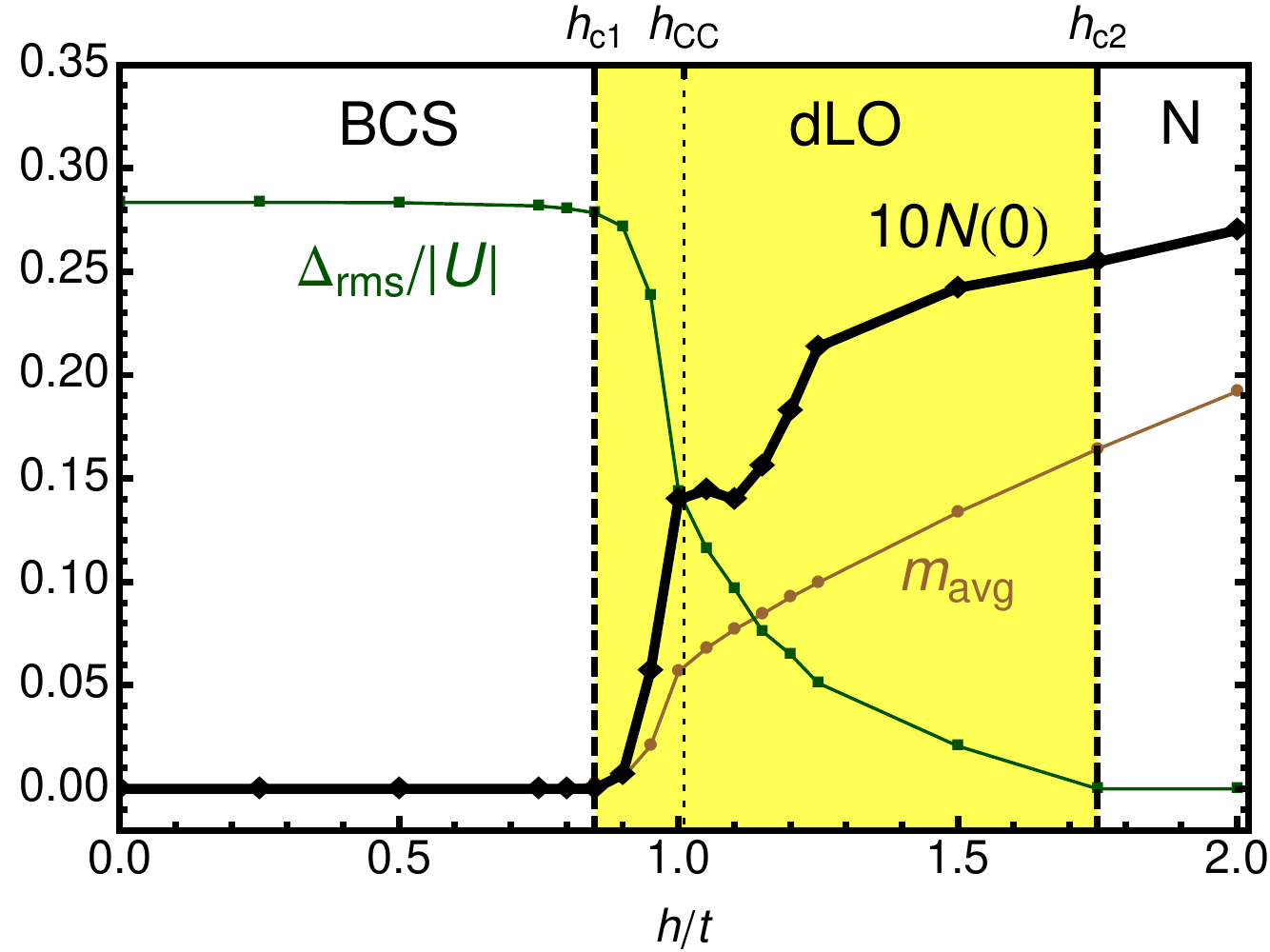}
   \caption{\small{
       Root-mean-square pairing amplitude $\Delta_\text{rms}$,
       average magnetization $m_\text{avg}$,
       and Fermi-level density of states $N(0)$
       as functions of Zeeman field $h$, 
       in units of the hopping amplitude $t$ (see \eref{hamiltonian}).
       For $h_{c1} < h < h_{c2}$ there is a disordered LO state
       with coexistent pairing and magnetization,
       in which the gap is partially filled in.
       The results are obtained using BdG simulations
       on a $36\times 36$ Hubbard model
       at weak disorder $W=1t$ (well below the critical disorder\cite{trivedi1996} for the destruction of superconductivity $W_c\sim 3t$), nonzero chemical potential $\mu=-0.25t$ to avoid perfect nesting effects at half-filling, low temperature $T=0.1t$, and a relatively large attraction $\left|U\right|=4t$ so that the coherence length is less than the system size.
			$h=\frac{1}{2} g\mu_B H$, where $g\approx 2$ is the $g$-factor, $\mu_B$ is the Bohr magneton, and $H$ is the parallel field.
       \label{BdGFrmsMavg}
       \vspace{-0.5cm}
   }}
   \end{figure}
\mysection{Experimental setup:} 
In the present study planar tunnel junctions formed on 3 nm-thick Al films were used to extract the low temperature quasiparticle DoS.  Aluminum has a well documented low spin-orbit scattering rate 
\cite{meservey1975}
and superconducting transition temperature $T_c=2.7$ K with a zero field gap $\Delta_o\approx 0.43$ mV in thin film form.  [For sample preparation see supplement].  
Measurements of
resistance and tunneling were carried out on an Oxford dilution
refrigerator using a standard {\it dc} four-probe technique.  Magnetic
fields of up to 9~T were applied using a superconducting solenoid. A
mechanical rotator was employed to orient the sample \textit{in
situ} with a precision of $\sim0.1^{\circ}$.
The films were moderately disordered with sheet resistances of the order of $1 k\Omega$,
well below the quantum of resistance for superconductivity 
$R_Q = h/{4e^2} = 6.4\ \mathrm{k\Omega}$.

  \begin{figure*}[!htb]
   \includegraphics[width=0.8\textwidth]{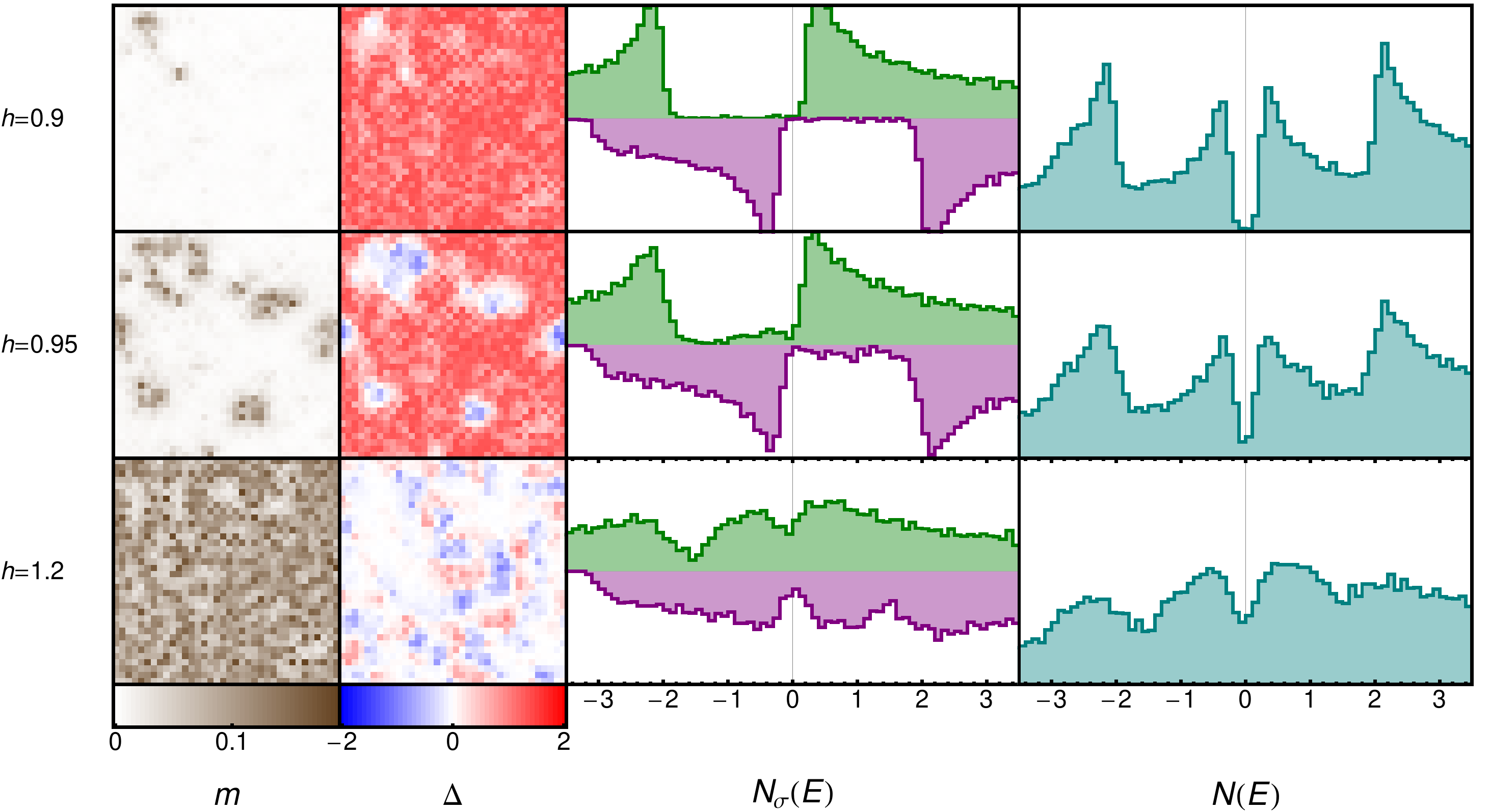}
   \caption{
       \small{The first two columns show spatial maps of
       the local pairing amplitude $\Delta$
       and the magnetization $m$.
       The third column show
       the densities of states (DoS's) of up and down electrons $N_\sigma(E)$.
       The last column shows the total DoS $N(E)$.
       For intermediate fields (e.g., $h/t=0.95$ and $h/t=1.2$)
       the system exhibits disordered Larkin-Ovchinnikov states
       with domain walls
       at which $m$ is finite,
       $\Delta$ changes sign,
       and the DoS becomes finite at low energy.
     Other parameters are as in Fig.~\ref{BdGFrmsMavg}.
       \label{EvolutionOfDoSWithZeemanField}
   }}
   \end{figure*}

\mysection{Experimental results and comparison with standard BCS theory:} 
\providecommand{\Tesla}{~\mathrm{T}}
We present measurements of the tunneling conductance $G$ of Al films,
	which is mainly proportional to the superconducting DoS
	at the low temperatures used.
Figure~\ref{GVdata} shows the bias dependence $G(V)$ in a parallel field $H=4.75 \Tesla$ at $100 ~\mathrm{mK}$,
	in which the BCS coherence peaks have been Zeeman-split by the applied field.   
Figure~\ref{GHdata} shows the parallel-field dependence of the zero-bias tunneling conductance $G(0)$,
	which is zero in the conventional superconducting state ($H < H_0 \approx 2.8 \Tesla$)
	and constant in the normal state ($H > H_{c\parallel} \approx 6.1 \Tesla$);
	however, there is a significant tail in $G(0)$ over a range of fields $H_0 < H< H_{c\parallel}$.
The colored curves in Fig.~\ref{GVdata} and \ref{GHdata} are obtained within
homogenous BCS mean field theory by solving the Usadel equations for the disorder-averaged
semiclassical Green's functions together with the self-consistent
equations for the \textit{uniform} order parameter and the internal magnetic field.
The parameters involved are the gap energy,
spin-orbit scattering rate, the orbital depairing rate, and the
antisymmetric Fermi-liquid parameter;
they are determined by fits\cite{Catelani2008,catelani:054512} to full spectra as in Fig.~\ref{GVdata}.

The observed excess zero-bias conductance $G(0)$ can have various origins.  
(i) Imperfect alignment: The inset of Fig.~\ref{GHdata},  shows $G(0)$  at several alignment angles between the film plane and the applied field.  It is evident that our alignment mechanism is precise enough to find parallel orientation within the limits of the sensitivity of the tunneling conductance to $H_\perp$, the perpendicular field component.   
(ii) Junction leakage is ruled out because all of the junctions used in this study had a very low zero-bias conductance in zero field, $G(2\mathrm{mV})/G(0) \sim 10^2-10^3$ at 100 mK.  
(iii) Material inhomogeneities: In principle could lead to broadened transitions, however, the zero-field gap in Al (and hence the nominal critical field $h_{CC}$) varies by only $20\percent$ over a very wide range of sheet resistance\cite{wu1994}  and averaging over a distribution of gaps fails to explain the large range of $H_\parallel$ over which $G(0)$ is finite.
(iv) Pair-breaking:  These effects scale as $Dd^3$, where $D$ is the normal state diffusivity and $d$ is the film thickness.    
For our films as $d$ is decreased from  3 nm to 2 nm, $D$ decreases by an order of magnitude, but $G(0)$ hardly changes.
Furthermore, recent tunneling measurements of Al-EuS bilayers have shown that a comparable $G(0)$  is produced by an interface-induced exchange field, which is a pure Zeeman field with no orbital depairing effects.
\cite{xiong2011}

 \begin{figure*}[!htb]
	\subfigure[]{
		\label{CombinedMap}
		\includegraphics[height=0.21\textwidth]{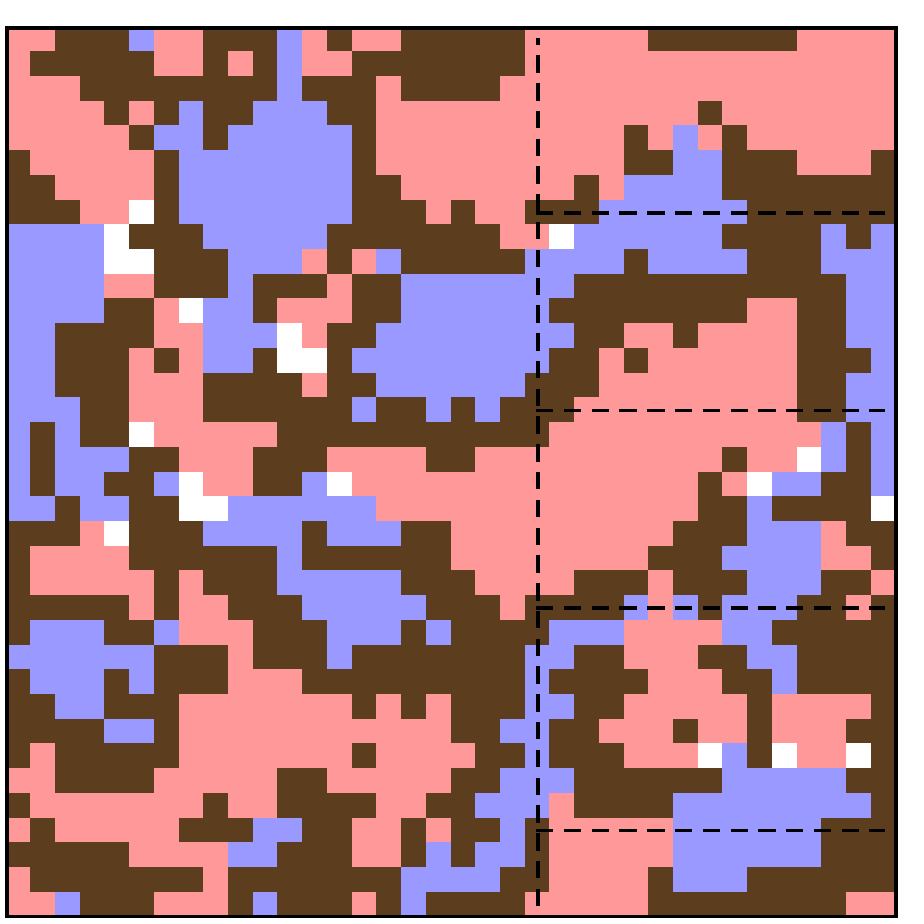}
		\hspace{-5mm}
	}
	\subfigure[]{
		\label{SlicePairing}
		\includegraphics[height=0.21\textwidth]{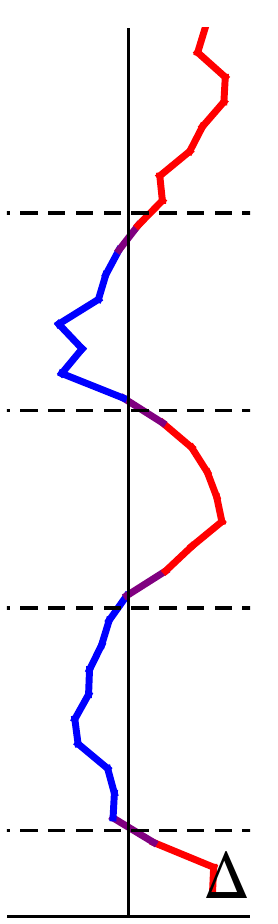}
		\hspace{-5mm}
	}
	\subfigure[]{
		\label{SliceMagnetization}
		\includegraphics[height=0.21\textwidth]{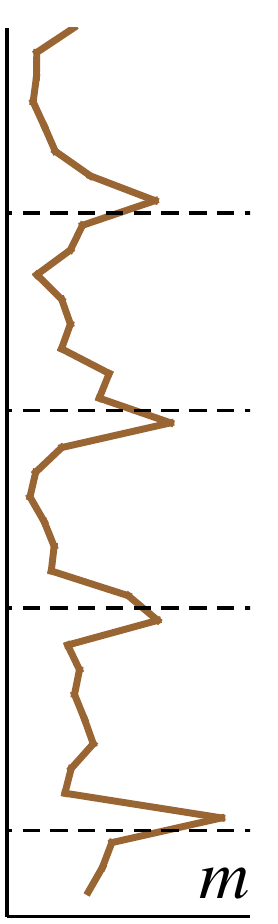}
	}
	\subfigure[$m(\rrr)$]{
		\label{MagnetizationMap}
		\includegraphics[height=0.21\textwidth]{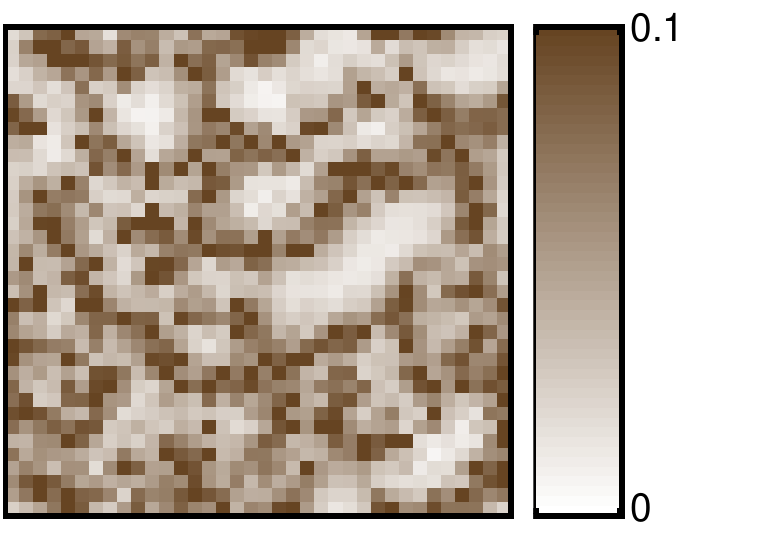}
	}
	\subfigure[$I(\rrr)$]{
		\label{LowEnergySpectralWeightMap}
		\includegraphics[height=0.21\textwidth]{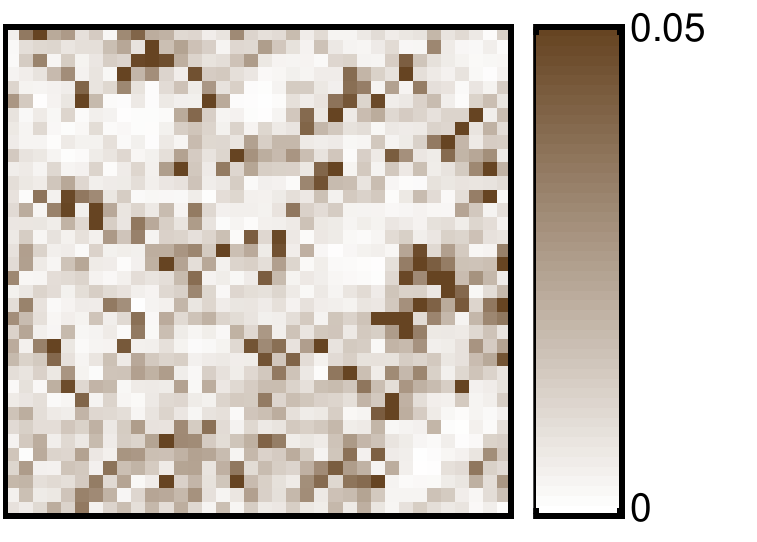}
	}
	\caption{
	\small{
	(a) Combined plot of $m(\rrr)$ and $\Delta(\rrr)$ for $h/t=1$ (other parameters as in Fig.~\ref{BdGFrmsMavg}).  
	Red (blue) indicates regions where $\Delta(\rrr)$ is large and positive (negative).
	Brown regions, where the magnetization $m(\rrr)$ is large,
		occur at domain walls where $\Delta$ changes sign.
	White regions are hills or valleys of the disorder potential
		corresponding to empty sites or localized pairs
		that participate in neither superconductivity nor magnetism.
	(b) and (c) show oscillations of $\Delta$ 
		along the vertical dashed line in panel (a).
	(d) and (e) show the correspondence between magnetization $m(\rrr)$
		and low-energy spectral weight $I(\rrr) = \int_{-0.1t}^{0.1t} dE~ N_\rrr(E)$.
	}
	}
	\label{LOphysics}
	\end{figure*}

\mysection{Disordered LO states and excess low-energy spectral weight:} 
Having ruled out all the above explanations,
we now argue that the anomalous excess zero-bias conductance at intermediate fields 
is an {\emph{intrinsic}} property of the condensate
due to the development of an exotic disordered Larkin-Ovchinnikov (dLO) phase
with an \emph{inhomogeneous} pairing amplitude and magnetization.

Our model consists of the attractive Hubbard Hamiltonian with a disorder potential and a Zeeman field,
 \begin{align}
 H &=
   \sum_{\rrr\rrr'\sigma} t_{\rrr\rrr'} \cdag_{\rrr\sigma} \cccc_{\rrr'\sigma}
   + \sum_{\rrr\sigma} (V_\rrr - \mu - h\sigma) (n_{\rrr\sigma} - \half)
   \nonumber\\&{}~~~
 - \left| U \right| \sum_{\rrr} (n_{\rrr\up} - \half)(n_{\rrr\dn} - \half)
 \elabel{hamiltonian}
 \end{align}
where
$t_{\rrr\rrr'}$ are hopping amplitudes (equal to $t$, taken as the unit of energy) between nearest-neighbor sites $\rrr$ and $\rrr'$,
$ n_{\rrr\sigma} = \cdag_{\rrr\sigma} \cccc_{\rrr\sigma} $
is the number operator for fermions of spin index $\sigma=\pm 1$ at site $\rrr$,
$\mu$ is the average chemical potential, $h$ is the Zeeman field,
and $U$ is the local pairwise Hubbard interaction.
The disorder potential $V_\rrr$ at each site is picked independently from a uniform distribution on $[-\frac{W}{2}, \frac{W}{2}]$.
We calculate the local densities $n_{\rrr\sigma}$, pairing amplitude 
$\Delta_{\rrr}=\left| U \right| \mean{c_{\rrr\dn} c_{\rrr\up}}$, and spin-dependent DoS $N_\sigma(E)$ within a fully self-consistent Bogoliubov-de Gennes (BdG) framework including all Hartree shifts (see supplement for details).  A phase diagram for this system was obtained in Ref.~\onlinecite{cui:054501}; in this paper we focus on spectral features.

As illustrated in Fig.~\ref{BdGFrmsMavg}, if $\Delta$ is restricted to be uniform, we find that the BCS- and normal-state free energies cross at $h_{CC}=1.01t$, the critical field for the 
first-order Chandrasekhar-Clogston transition (here $h_{CC}$ differs from $\Delta_0/\sqrt{2}$ due to the moderate value of $U$).
However, if $\Delta(\rrr)$ is allowed to be inhomogeneous, BdG calculations predict \emph{two} transitions, 
at a lower critical field $h_{c1}=0.85t$ and an upper critical field $h_{c2}=1.75t$.
The intermediate state ($h_{c1} < h_{CC} < h_{c2}$) has both a finite pairing amplitude and a finite magnetization.

A physical understanding is provided in Fig.~\ref{EvolutionOfDoSWithZeemanField}, which shows the local pairing amplitude $\Delta(\rrr)$, local magnetization $m(\rrr) = \half \left[ n_\up(\rrr) - n_\dn(\rrr) \right]$, and spatially averaged DoS's of up and down spins $N_\sigma(E)$, for various values of $h$.
At low fields the system is a BCS superconductor with a nearly uniform order parameter $\Delta(\rrr) \approx \Delta_0$, 
whose DoS contains coherence peaks at $\pm \Delta \pm h$ slightly broadened by inhomogeneous Hartree shifts\cite{ghosal1998,ghosal2001}.
At high fields the system is normal (non-superconducting) with nearly uniform magnetization.
At intermediate fields there is a disordered Larkin-Ovchinnikov (dLO) state with the following features:
(1) There is a strong modulation of the pairing amplitude $\Delta(\rrr)$ which changes sign between positive and negative values.
The oscillations at wavevector $q_\text{LO} \approx 2k_F$ are partially disrupted by the disorder potential.
(2) The magnetization is finite in the domain walls where the pairing amplitude is small.
(3) \emph{There is significant low-energy weight in the DoS}, as illustrated in the rightmost column of Fig.~\ref{EvolutionOfDoSWithZeemanField}.  This is the main new result of this paper, and it is a likely explanation for the similar low-energy weight seen in experiments (Fig.~\ref{Gdata}).

\mysection{Origin of low-energy states:} 
When the Zeeman field exceeds a certain lower critical field, 
magnetization begins to penetrate the sample in the form of domain walls
(brown regions in Fig.~\ref{CombinedMap}).
The majority electrons are unable to enter the superconducting regions due to the gap, and so they are confined to the domain walls by Andreev reflection, forming Andreev bound states with a distribution of energies.  Whereas in a clean LO state\cite{bakhtiari2008,loh2010} tunneling between domain walls gives rise to subgap bands, in a dLO state the bound states are likely to remain localized, but they still contribute to the low-energy DoS.  Indeed, comparing Figs.~\ref{LOphysics}(d) and (e) shows that the low-energy weight is concentrated in the same regions as the magnetization. 
The tunneling DoS (unlike transport measurements) is sensitive to local electronic structure, and hence the low-energy spectral signatures of LO should remain even when phase fluctuations prevent the development of long-range LO order.\cite{radzihovsky2009}

We conclude that dLO physics is a likely explanation of the longstanding mystery of excess zero-bias tunneling conductance of Al films near the spin-paramagnetic transition.\cite{adams2004}
Our results suggest that the parallel-field-tuned\cite{zhou1998,dubi2007} 
superconductor-insulator transition (SIT) occurs via a dLO phase in which the gap is \emph{filled in} by Andreev bound states.
This scenario is distinct from the zero-field thickness-tuned ``fermionic'' SIT where the gap \emph{closes}\cite{gantmakherreview,valles1992,finkelstein1994}, and from the ``bosonic'' SIT~\cite{fisher1989,ghosal1998,ghosal2001,bouadim:arXiv,nguyen2009} 
where the gap appears to \emph{remain finite} across the SIT.

We acknowledge support from the U.S. Department of Energy, Office of
Basic Energy Sciences, Division of Materials Sciences and Engineering
under Awards DE-FG02-07ER46423 (YLL,NT) and DE-FG02-07ER46420 (PWA).
GC is supported by Yale University.



\end{document}


\title{Mystery of Excess Low Energy States \\ in a Disordered Superconductor
in a Zeeman Field: \\
Supplementary Information}
\maketitle

\section{Sample Preparation}
In the present study tunnel junctions were formed by first depositing a 3 nm thick Al film e-beam deposition of 99.999\% Al stock onto
fire polished glass microscope slides held at 84~K.  After deposition, the film was exposed to the atmosphere for 10-20 minutes in order to allow a thin native oxide layer to form. Then a non-superconducting Al counterelectrode was deposited from an Al 2024 alloy target, with the oxide serving as the tunneling barrier.  The low temperature parallel critical fields of the counter-electrodes were $\sim$6~T, in good agreement with the expected $H_{c||}$. The junction area was about 1~mm$\times$1~mm, while the junction resistance ranged from 15-100~k$\Omega$ depending on exposure time and other factors.  Only
junctions with resistances much greater than that of the films were used, in order to be in the tunneling regime.


\section{Variational Bogoliubov-de Gennes Method}
The combination of the Zeeman field and the disorder potential ultimately leads to inhomogeneous, spin-dependent Hartree potentials.  Therefore, we use a  generalized Bogoliubov-de Gennes (BdG) method\cite{degennes} in which all $2N$ BdG eigenvalues and eigenvectors are distinct (where $N$ is the number of sites).

For convenience, we write the Hamiltonian in terms of an applied chemical potential $\mu_\rrr=\mu - V_\rrr$ (where $V_\rrr$ is the quenched random potential) and field $h_\rrr=h$ at every site:
 \begin{align}
 H &=
   \sum_{\rrr\rrr'\sigma} t_{\rrr\rrr'} \cdag_{\rrr\sigma} \cccc_{\rrr'\sigma}
   - \sum_{\rrr\sigma} (\mu_\rrr + h_\rrr \sigma) x_{\rrr\sigma}
 +U \sum_{\rrr} x_{\rrr\up} x_{\rrr\dn}
	,
 \end{align}
where $x_{\rrr\sigma} = n_{\rrr\sigma} - \half$ are densities with respect to half-filling and $U<0$ represents attraction.  

We decouple the Hubbard interaction in charge, spin, and pairing channels.
It is difficult to justify a traditional derivation of the self-consistent BdG equations with multiple-channel decoupling, because this appears to overcount the interaction term.
We have performed a rigorous derivation based on the $\Tr \rho \ln \rho$ variational formalism.\cite{chaikin} 
In this approach, we postulate a trial Hamiltonian $\hat{H}_\text{trial}$, which defines a trial density matrix 
$\hat{\rho}_\text{trial} \propto \exp (-\beta\hat{H}_\text{trial})$,
and we then minimize the variational free energy $\Omega$ [given in \eref{VFE}]
with respect to the $3N$ variational parameters: the Hartree chemical potential $\mu^H_\rrr$, Hartree field $h^H_\rrr$, and self-consistent pairing field $\Delta_\rrr$.
This formalism has the practical advantage that $\Omega$ can be used to assess the quality of the variational approximation \emph{during} the approach to self-consistency, and that it provides a rigorous upper bound to the true free energy.
Our implementation is as follows:

\begin{enumerate}
\item
Make arbitrary initial guesses for the Hartree chemical potential $\mu^H_\rrr$, Hartree field $h^H_\rrr$, and self-consistent pairing field $\Delta_\rrr$ at every site $\rrr$.  These constitute a set of $3N$ real-valued variational parameters.

\item
Find the effective chemical potential $\mutot_\rrr = \mu_\rrr + \mu^H_\rrr$ 
and effective field $\htot_\rrr = h_\rrr + h^H_\rrr$
at every site.  These effective potentials include both the applied potentials and the Hartree potentials (resulting from the decoupling of the $U$ term); they enter the mean-field Hamiltonian,
 \begin{align}
 H &=
   \sum_{\rrr\rrr'\sigma} t_{\rrr\rrr'} \cdag_{\rrr\sigma} \cccc_{\rrr'\sigma}
   - \sum_{\rrr\sigma} (\mutot_\rrr + \htot_\rrr \sigma) x_{\rrr\sigma}
	.
 \end{align}

\item
Construct the $2N\times 2N$ Hamiltonian matrix $H_{\rrr\sigma;\rrr'\sigma'}$, where the indices $\sigma,\sigma'=\up,\dn$ distinguish between up-particle and down-\emph{hole} sectors connected by matrix elements $\Delta$:
   \begin{align}
   H_{\rrr\rrr' \sigma\sigma'}
   &=
   -t_{\rrr\rrr'} \pmat{1 & 0 \\ 0 & -1}_{\sigma\sigma'}
   \nonumber\\&~~{}
   -\delta_{\rrr\rrr'}  \pmat{\mutot_\rrr+\htot_\rrr & \Delta_\rrr \\ \bar\Delta_\rrr & -\mutot_\rrr+\htot_\rrr}_{\sigma\sigma'}
	.
   \end{align}

\item
Diagonalize $\mathbf{H}$ to obtain eigenvalues $E_\alpha$ and eigenvectors $\phi_{\alpha \rrr \sigma}$, where the eigenmode index $\alpha$ runs from $1$ to $2N$.
(These eigenvectors are generalizations of the $u_{\alpha\rrr}$ and $v_{\alpha\rrr}$ vectors that appear in the original BdG formalism.)

\item
Find the symmetrized occupation numbers $\zeta_\alpha = -\half\tanh \half\beta E_\alpha$.

\item
Compute the number densities $x_{\rrr\sigma}$ (relative to half-filling) and the pairing density $F_{\rrr}=\mean{\cccc_{\rrr\dn} \cccc_{\rrr\up}}$ at every site $\rrr$:
   \begin{align}
   x_{\rrr\up} &=  \sum_\alpha \zeta_\alpha 
   	\phi^*_{\alpha\rrr \up} \phi_{\alpha\rrr \up}
,\\
   x_{\rrr\dn} &= -\sum_\alpha \zeta_\alpha 
   	\phi^*_{\alpha\rrr \dn} \phi_{\alpha\rrr \dn}
,\\
   F_\rrr &= \sum_\alpha \zeta_\alpha 
   \left( \phi^*_{\alpha\rrr \up} \phi_{\alpha\rrr \dn} + h.c. \right),
   \end{align}
and thence the number density and magnetization on each site,
   \begin{align}
   x_{\rrr} &= \half \left( x_{\rrr\up} + x_{\rrr\dn}  \right)
,\\
   m_{\rrr} &= \half \left( x_{\rrr\up} - x_{\rrr\dn}  \right)
.
   \end{align}

\item
Compute the variational free energy
   \begin{align}
   \Omega &=
   \sum_\alpha \ln (2 \cosh \half\beta E_\alpha)
\nonumber\\&~~{}
   + \sum_\rrr U(F_\rrr^2 + x_\rrr^2 - m_\rrr^2)
\nonumber\\&~~{}
   + \sum_\rrr 2(\Delta_\rrr F_\rrr + \mu^H_\rrr x_\rrr + h^H_\rrr m_\rrr )
.
	\elabel{VFE}
   \end{align}

\item
According to the usual variational principle, we wish to minimize $\Omega$ with respect to $\Delta$, $\mu^H$, and $h^H$ (to obtain a least upper bound to the true free energy).  In practice this can be done by solving the stationarity condition $\nabla\Omega=0$, i.e., finding a root of the $3N$-dimensional equation
	\begin{align}
		\fff (\XXX) = \mathbf{0},
	\elabel{residual}
	\end{align}
where $\XXX = \{ \mu^H_\rrr, h^H_\rrr, \Delta\rrr  \}$ is the vector of variational parameters 
and 
	\begin{align}
		\fff = \{ \mu^H_\rrr+U x_\rrr,   ~   h^H_\rrr-U m_\rrr,   ~  \Delta_\rrr+U F_\rrr  \}
	\elabel{residualdefn}
	\end{align}
is the residual vector (the ``distance'' from self-consistency).
We use the standard Broyden method,\cite{numericalrecipes} which is a superlinearly convergent quasi-Newton method for multidimensional root-finding.
The first iteration of the Broyden procedure is equivalent to fixed-point iteration of the self-consistency equations
   \begin{align}
   \Delta^H_\rrr &= -U F_\rrr
,\\
   \mu^H_\rrr &= -U x_\rrr
,\\
   h^H_\rrr &= +U m_\rrr
.
	\elabel{FPI}
   \end{align}
We also inspect $\Omega$ to verify that the root of \eref{residual} corresponds to a minimum of \eref{VFE}, and not to a maximum.  We restart the Broyden method using \eref{FPI} if a Broyden step results in a large increase in $\Omega$ (since quasi-Newton methods are prone to instability).   
   
\end{enumerate}

After convergence we calculate further quantities, including the densities of states for up and down electrons (which are the main point of interest in this paper):
   \begin{align}
   N_{\rrr\up} (E)   &= \sum_\alpha \delta(E - E_\alpha)      
    \phi^*_{\alpha\rrr \up} \phi_{\alpha\rrr \up}  , \nonumber\\
   N_{\rrr\dn} (E)   &= \sum_\alpha \delta(E + E_\alpha)       
   	\phi^*_{\alpha\rrr \dn} \phi_{\alpha\rrr \dn}  .
   \end{align}

